Babinet's principle in the Fresnel regime studied using ultrasound

Akira Hitachi[a)] and Momo Takata
Kochi Medical School, Nankoku, Kochi 783-8505, Japan.

The diffraction of ultrasound by a circular disk and an aperture of the same size have been investigated as a demonstration of Babinet's principle in the Fresnel regime. The amplitude and the phase of diffracted ultrasonic waves have been measured and a graphical treatment of the results is performed. The results verify Babinet's principle. It is also found that the incident wave is indeed $\pi/2$ behind the phase of the wave passing through on the central axis of a circular aperture. This paradox has previously been regarded as a defect of Fresnel's theory.

## I. INTRODUCTION

The high intensity spot just behind a circular opaque obstacle, known as the Poisson's bright spot (the spot of Arago),[1,2] arouses curiosity and can be used to gain the attention and interest not only of students of physics but also of those of general science classes. The additional observation of the complementary circular aperture provides materials for quantitative discussions on the mathematical theory of waves.

A general description of Babinet's principle states that the sum of the diffraction fields $U_1$ and $U_2$ behind two complementary objects is the field $U_0$ observed without the diffracting objects.[3-8]

$$U_0 = U_1 + U_2 \ . \tag{1}$$

This principle is generally fulfilled within the framework of the scalar theory of diffraction.[8] Verification of the Babinet's principle has been almost exclusively performed with light. The diffracted intensity patterns ($I = UU^*$) are compared in the Fraunhofer regime[4,7] (source and observation plane situated very far from the diffracting object). In the Fresnel regime (source and observation plane at a finite distance from the diffracting object), the intensity patterns are different from each other and are compared through calculation.[1] This is mainly because in optics the wavelength is too short and the frequency too high to observe the phase. In the microwave region, the wavelength becomes a few cm. It is possible to make use of this advantage, e.g. the phase dependence of the dielectric transient absorption allows observation of a phenomenon several orders of magnitude faster than the time resolution which is determined by the $Q$ value of the cavity.[9] The wavelength of ultrasound is also several mm and the period is a few tens of microseconds. Then, it is possible to manipulate and observe the phase as well as the amplitude. This enables Babinet's principle to be studied directly by comparing the amplitude and the phase. With ultrasound measurements the essence of the theory can be seen by simply drawing vectors (phasors) in the complex plane, without complicated calculation.

We will observe diffraction by a circular aperture having a diffracted field $U_a$ and a complementary circular disk $U_\mathrm{d}$ and study Babinet's principle, $U_0 = U_a + U_\mathrm{d}$, by using ultrasound. The apparatus is inexpensive and can be assembled by hand. The measurement is as simple as learning how to use the oscilloscope. A graphical treatment, the vibration spiral, based on Huygens-Fresnel diffraction theory, is used to analyze the observed amplitude and phase.

A curious feature of both the graphical approaches and mathematical analysis is a $\pi/2$ advance of the phase for the wave passing through on the central axis of a circular aperture with respect to the incident wave $U_0$. This phase difference has been regarded as a defect of Huygens-Fresnel diffraction theory. However, the phase difference has never been actually measured as far as we know. We will measure the phase due to a small aperture with a diameter close to the wavelength being used. The vibration spiral for the circular aperture will be investigated in detail.

## II. THEORETICAL BACKGROUND

Huygens' principle explains the propagation of a wave as follows: every point of a wave-front may be regarded as a source of secondary spherical wavelets, and the wave-front at any later instant is constructed as the envelope of these wavelets as shown in Fig. 1.[5, 10] Fresnel constructed a diffraction theory based on Huygens' principle together with the principle of interference. We will discuss the theoretical basis within two alternative approaches of a graphical and mathematical treatments.

### A. Graphical treatment of diffraction by a circular object

A graphical treatment, based on Huygens-Fresnel diffraction theory, gives a clear physical picture of the origin of the diffraction pattern.[4,6,7] Consider a point source *S* located at a distance $z_0$ from a circular aperture and the diffracted field $U_a$ at point *P* on the axis at a distance $z$ from the aperture. Consider now a set of imagined spheres centered at *P* of radii $z + \lambda/2$, $z + 2\,\lambda/2$, $z + 3\lambda/2$ and so on, differing by $\lambda/2$ as shown in Fig. 1. Their intersections at the aperture with a spherical wave front emanating from the source S form circular zones, $F_1$, $F_2$, $F_3$ and so on. The zones are called Fresnel or half-period zones. There is a $\pi$ phase difference between consecutive zones, and therefore, the contributions of the successive zones to the diffracted field are alternately positive and negative. Figure 2(a) shows the amplitude diagram in the complex plane when the first half-period zone is divided into eight subzones.[4] The phasor $a_1$ represents the contribution from the first subzone. The phasors $a_2$ and $a_3$ due to the second and third subzones are added. The phase difference $\delta$ between each successive zone will be $\pi/8$. Adding all eight subzones gives the phasor *OG* as the resultant amplitude from the first half-period zone. Similarly, the contribution from the second half-period zone is *GH*, and *OH* is the sum of the first two zones. Successive half-period zones give the rest of Fig. 2(a). By increasing the number of subzones indefinitely, the vibration spiral is obtained as shown in Fig. 2(b). $U_a$ is represented by phasor *OD*. The spiral eventually approaches $C$; the sum of all the half-period zones covers the whole spherical wave, which should be the phasor $OC$ representing the unobstructed field $U_0$.

In Babinet' principle, $U_0 = U_a + U_d$, where $U_d$ is $DC$. The spiral is practically circular for many turns. Huygens-Fresnel diffraction theory predicts that the Poisson's bright spot appears behind a circular disc, its amplitude being almost equal to $U_0$. The argument here also applies for the plane wave geometry.

The spiral also allows one to determine directly the resultant amplitude and the phase due to any fractional number of zones. The phase can be estimated by simply calculating $\Delta_B$, the difference between the path length along the axis and that via the edge of the circular object. The angle φ ($\angle OCD$) on the spiral (Fig. 2) is given by $2\pi\times\Delta_B/\lambda$. Since $\Delta_B$ can be larger than λ, φ can exceed 2π.

Figure 2 applies for the time $t = 0$. The phasors revolve clockwise with the same angular velocity ω. The triangle representing $U_0$, $U_a$ and $U_d$ revolves as a rigid frame. The simple harmonic wave is presented as the projection on the coordinate of the real part. We adopt here the custom in optics of representing an advance of phase by a clockwise rotation of the amplitude phasor.

A problem arises. The resultant phasor $U_0$ turns out to be π/2 behind the wave from the center of the zone system, whereas, according with Huygen's principle, a wavelet emitted at $O$ (Fig. 1) in phase with the incident wave $U_0$ should arrive at $P$ still in phase with $U_0$. This paradox has been regarded as a defect of Fresnel's theory resulting from the approximations made therein.[4,7]

B. Mathematical treatment in relation to the vibration spiral

We will discuss now the scalar theory of diffraction in relation to the vibration spiral. Kirchhoff gave a mathematical expression for Fresnel's wave theory on the basis of Green's theorem. Fresnel-Kirchhoff diffraction theory provides an accurate solution in many instances and used most widely. The diffraction patterns for the circular aperture and disk are given by Lommel functions,[1] which are described in terms of the Bessel functions. We consider for simplicity the diffraction on the axis.

Fresnel-Kirchhoff diffraction for a point source and a circular aperture in polar coordinates can be expressed in terms of the incident wave $U_0(P)$. One obtains for point $P$ on the axis,[6]

$$U_a(P) = -iU_0(P)\int_0^{\varphi_0} e^{i\varphi} Q(\varphi)d\varphi\,, \tag{2}$$

$$U_0(P) = A_0 \exp\{-i(\omega t - kr - \theta)\}/r\,, \tag{3}$$

where $k = 2\pi/\lambda$, $A_0$ is the amplitude of the incident wave, $\varphi_0$ is the value of the phase difference at $P$ between a wave that comes from the edge of the aperture and that which follows the $z$-axis. The inclination factor $Q(\varphi) = ½(\cos\alpha + \cos\beta)$ is a smooth monotonically decreasing function. $\alpha$ and $\beta$ are angles that the diffracted and incident waves, respectively, would make with a normal unit vector **n** perpendicular to the diffraction plane (Fig. 1). The $-i$ appearing in Eq. (2) accounts for a $\pi/2$ phase shift of the diffracted wave relative to the incident wave.[5-7,10] The phasor integral

$$I(\varphi_0) = \int_0^{\varphi_0} Q(\varphi)e^{i\varphi}d\varphi \tag{4}$$

gives the vibration spiral.[6] For a few turns $Q(\varphi)$ is close to 1 . Therefore, the spiral may be regarded as a circle. $I(\varphi_0) = 1+i$, $2i$, $-1+i$ and 0 for the phase difference $\varphi_0 = \pi/2$, $\pi$, $3\pi/2$ and $2\pi$, respectively, and the points for $I(\varphi_0)$, multiplied by $A_0$, will lie on the circle in Fig. 2(b). In the limit of infinite aperture radius, $I(\varphi_0)$ becomes $i$, and $U_a(P)$ approaches $U_0(P)$, the incident wave. The resultant phase from Eq. (2) is $-\pi/2$ for a wave that follows the $z$ axis and 0 for no aperture. The vibration spiral in Fig. 2 should be turned $\pi/2$ clockwise, then $U_0$ is on the real axis.

However, Fresnel-Kirchhoff diffraction presents inconsistencies in the boundary conditions and is valid when the diffraction object is large compared with the wavelength ($D>>\lambda$) but small compared with $z_0$ and $z$ ($z_0$, $z >> D$). [3,5]

Rayleigh-Sommerfeld diffraction theory gives a rigorous expression for the scalar field.[11] This theory has not been discussed in relation to the vibration spiral. Closed solutions for the diffracted field for the spherical wave by a circular aperture and a disk along the axis is given as,[12]

$$U_a(z) = A_0\left[\frac{\exp\{ik(z_0+z)\}}{(z_0+z)} + \exp(i\pi)\frac{z}{Z}\frac{\exp\{ik(Z_0+Z)\}}{(Z_0+Z)}\right], \tag{5}$$

$$U_d(z) = A_0\frac{z}{Z}\frac{\exp\{ik(Z_0+Z)\}}{(Z_0+Z)}, \tag{6}$$

where $Z_0$ is the distance between the edge and the point source on the axis and $Z$ is the distance between the edge and the observation point on the axis (Fig. 3). These simple formulas are valid anywhere along the axis behind the object. The first term in Eq. (5) is the incident spherical wave $U_0(z)$ and the second term is $-U_d(z)$, ($\exp(i\pi) = -1$). We have the Babinet's principle, $U_0 = U_a + U_d$, on the axis.[12]

Now we move into the Fresnel regime ($z_0$, $z > D$) and look at $U_d(z)$ as a function of $D$ at a fixed $z$ on the axis. Eq. (6) is rewritten as,

$$U_d(z) = A_0 \frac{z}{Z}\frac{z_0+z}{Z_0+Z}\frac{\exp\{ik(Z_0+Z)\}}{(z_0+z)}. \tag{7}$$

The amplitude $A_d$ for a disk is given by $(z/Z)\cdot(z_0+z)/(Z_0+Z)\cdot A_0$ and is almost equal to $A_0$ when $D$ is not too large. The phase difference from the incident wave is given by $\varphi = 2\pi\Delta_B/\lambda = 2\pi\{(Z_0+Z)-(z_0+z)\}/\lambda$. Let us put the endpoint of phasor $U_d$ onto that of $U_0$ and see the trace of the starting point of $U_d$, which is also the end point of $U_a$, on the complex plane by increasing $D$. Then we have the vibration spiral as shown in Fig. 2(b), except $U_0$ is on the real axis. For a very small aperture, $\Delta_B$ is small and $U_d$ becomes close to $U_0$. Therefore, the resultant phasor $U_a = U_0 - U_d$ has a phase $\pi/2$ ahead from $U_0$. Fresnel zones can be obtained as $(z_0+z)+\Delta_B = (z_0+z)+n\cdot\lambda/2$, where $n$ =1 ,2, 3 and so on.

The diffracted field for the plane wave by a circular aperture along the axis is expressed simply,[13,14]

$$U_a(z) = A_0 z\left[\frac{\exp(ikz)}{z} - \frac{\exp(ikZ)}{Z}\right]. \tag{8}$$

An one-dimensional integral over the rim of the aperture[13] and a conventional two-dimensional integration over the area of the aperture[14] yield the same result.

As we have seen, the theories all show a $\pi/2$ phase difference between the incident wave and the wave passing through on the central axis of a circular aperture.

## III. EXPERIMENTAL METHOD

The experimental setup using ultrasound is shown schematically in Fig. 3. The transmitter $T$ (Nicera T4016) was driven at 40 kHz by an oscillator. This gives a wavelength $\lambda$ of 8.58 mm at room temperature. The frequency was monitored by a frequency counter. The transmitter and the center of the diffracting objects were placed on the axis so that the ultrasonic path has cylindrical geometry about the normal to the object through its center. The receiver $R$ (Nicera R4010A1) was chosen because it has a wide signal accepting angle (95 $^{o}$ FWHM) and was set on a rail placed perpendicular to the axis and scans the radial distribution of the signal beyond the circular aperture or a disk. The distances $z_0$ and $z$ are from the diffracting object (the aperture or the disk) to the source and the scanning line respectively. A transmitter with narrow emitting angle (50 °FWHM) was chosen in order to avoid undesirable reflection mainly from the table. The distance $z_0$ was set to be larger than $z$ in order to have a uniform sound pressure at the aperture plane. The signal from the receiver was put into ch1 of an oscilloscope and the voltage to the transmitter was monitored in ch2. The voltage used was typically 2 - 4 $V_{p\text{-}p}$. Then the amplitude of the received signal and the phase difference with respect to the signal in ch2 were observed. The amplitude and the phase were registered as a function of the radial distance *r*. The diameters of the transducers were 7 mm and 6 mm for T4016 and R4010A1, respectively.

A. Circular aperture and disk for study of Babinet's principle

A circular aperture and a complementary disk of 10 λ in diameter were prepared for the study of Babinet's principle. A circular aperture of 86 mm in diameter cut into 0.6 mm thick A4 size hard paper was placed on a cork board of 900 mm wide, 600 mm high and 6 mm thick with a circular hole of 118 mm in diameter. The cork board acts as a frame supporting the aperture and also absorbs the sound. A circular disk of the same diameter was also prepared. A circular disk taken from the same paper was backed by a circular cork disk of 69 mm in diameter and 4 mm thick. The disk was supported by a carbon rod of 4 mm in diameter. The apparatus was set up on a 5cm-square-ruled mat which covered the desk to guide alignment. The heights of the centers of the transmitter, the receiver, the aperture and the disk were set to be 340 mm.

The receiver was put on a rail placed 858 mm (100λ) away from the transmitter so that the radial distribution of the diffracted wave can be obtained. The position on the vibration spiral was selected by altering the position *z* of the aperture as shown in Table I.

B. Small aperture for phase study of vibration spiral

According to the vibration spiral, the phase difference Δθ observed on the axis for a circular aperture field $U_a$ relative to the unobstructed field $U_0$ will be –π/2 in the zero diameter limit, and -π/4, 0, and π/2, respectively, for an aperture containing a quarter-period zone, a half-period zone, and one period zone (Fig. 2). Given the relatively long wavelength of the ultrasound, the diameter of the first Fresnel half-period zone can be several cm. Therefore it is possible to observe effects within a small fraction of the half-period zone. A small aperture, which will contain only a tiny fraction of half-period zones, was prepared. The phase differences comparing with and without the aperture were measured.

We used a small circular aperture of 1.4λ in diameter and used the graphical treatment. A4 size hard paper of 0.6 mm thick with a 12mm diameter hole at the center was placed on the aperture system used in experiment A. The $z_0 + z$ distance was also set to be 100λ. The fraction of half-period zone covered was changed by altering the position of the small aperture. The aperture was set between $z$ = 17λ - 38λ.The fraction can be as small as 1/30 - 1/50 half-zone. The phase and the amplitude of the unobstructed wave $U_0$ on the axis were compared with the corresponding phase and amplitude of the wave $U_{sa}$ with the small aperture. The phase was also observed off the axis as a function of radial distance $r$ at $z$ = 17λ and 38λ. The position φ on the vibration spiral and the expected phase difference Δθ (= -90°+ φ/2) on the axis are shown in Table II. Sublabels 1, 2 and 3 stand for the distances $z$ of 17λ, 26.3λ and 38λ, respectively.

## IV. RESULTS

### A. Babinet's principle

The relationship between $U_0$, $U_a$ and $U_\mathrm{d}$ for the 10λ circular aperture and the disk placed 17λ, 26.3λ and 38λ, from the receiver set on the axis are shown in Fig. 4. The distance between the transmitter and the receiver was 100λ. The experimental parameters are shown in Table I. The end points for phasors $U_a$ and $U_0$ - $U_\mathrm{d}$ are plotted as open and closed symbols. Measurements of $U_a$ and $U_\mathrm{d}$ can be compared with theory independent of each other in this plot. The phase $\theta_0$ for no diffracting object was taken to be π/2 following the custom of representing the Fresnel vibration spiral. If the assumption in graphical treatment is true, the measured points lie on the spiral. The amplitude values for the aperture $A_a$ and the disk $A_\mathrm{d}$ are normalized to $A_0$. The φ values calculated by the path difference $\Delta_\mathrm{B}$ for each setup are shown with bars on the circle. The aperture contains less than two half-period zones with these setups. The scattered points are results for different measurements. The arrows show the average values for $z$ = 17λ. The main contribution to the error comes from errors in the phase. The amplitude for $U_\mathrm{d}$ is observed to be larger than that for $U_0$ and points are mostly outside the circle except for $z$ = 17λ. On the other hand, most points for $U_a$ lay inside of the circle. The reason may be due to the finite thickness of the edges, and the limited radial width of incident ultrasound. Although, there still are small discrepancies in points for $U_a$ and $U_\mathrm{d}$, the agreement between the graphical treatment and experimental results is good. It has also been shown that the point moves counter clockwise on the spiral as $\Delta_\mathrm{B}$ increases when the object distance is altered from the center to the receiver.

The $A_a$ values on the axis can be between 0 and twice $A_0$ depending on the position $z$ of the aperture. In the present setups, two $A_a$ maxima were observed at around $z = 44\lambda$ and $56\lambda$ where $\Delta_B$ is the same. $A_a$ was observed to be about 1.9 $A_0$ and $\theta_a$ was almost equal to $\theta_0$ at the maximum. An $A_a$ minimum was observed at $z = 14\lambda$. $\theta_a$ changed abruptly from being behind to ahead of $\theta_0$ as the aperture approaches the receiver in the vicinity of the minimum.

The amplitude and the phase measurements were also performed for off-axis positions. The typical radial diffraction patterns, the amplitude $A$ and the phase $\theta$, obtained for $U_0$, $U_a$ and $U_d$ are shown in Fig. 5. There are five characteristic positions shown in Fig. 5. One is the axis, and two are geometrical edges (vertical broken lines) which lay symmetric about the axis. Other positions $M_1$, where all phases come close, correspond to positions for the eclipse of the first half-period zone, (dot-dash lines). When the observation point $P$ moves out from the center and goes into the geometrical shadow of the aperture, the Fresnel zone system follows. As $P$ moves further, the total eclipse of a half-period zone occurs.

Phases also come close to each other at the geometrical edges. The position of the geometrical edge and the position where three phases coincide differ a little from each other. The difference may be due to the finite sizes of the transducers.

$A_0$ and $\theta_0$ change smoothly as expected over the whole range observed. Observed $A_0$ values are considerably flat and almost constant. $\theta_0$ lags as the receiver moves out from the axis and the distance from the source increases. The parabolic curve shows the phase shift of $U_0$, $2\pi\{TR'-(z_0+z)\}/\lambda$, calculated by the difference in transmitter-receiver distance between the receiver on the axis ($z_0$+$z$) and on a radial position ($TR'$). The curve represents the phase of a spherical wave originated at the transmitter observed on a line perpendicular to the axis as a function of radius $r$. Observed $\theta_0$ agree well with the calculated curve as expected.

The amplitude $A_a$ for the aperture increases slowly when the receiver approaches the geometrical edge from outside. Two $A_a$ maxima lay symmetric about the axis inside the geometrical edges. $A_a$ on the axis depends on $z$ as mentioned above. The amplitude $A_d$ for the disk is almost constant and the same as $A_0$, outside the shadow except a small minimum near $M_1$, until the receiver approaches the geometrical edge. Then it starts decreasing as the receiver comes close to the geometrical edge. Structures are seen in the shadow. Poisson's bright spot appears on the axis surrounded by small maxima. The amplitude of the spot is almost equal to $A_0$ as predicted by the graphical treatment.

Change in $\theta_a$ is slow inside the edge, except in the vicinity of the center. However, the phase $\theta_a$ changes rapidly outside the edge (in the geometrical shadow). Outside the edge, $\theta_a$ goes behind as $r$ increases until it approaches position $M_1$, then $\theta_a$ starts the reversal movement. The phase $\theta_a$ goes ahead instead of behind. After passing through $M_1$, $\theta_a$ again goes behind as $r$ increases. However, $A_a$ is small and the changes in $A_a$ are not evident in this region. The same behavior is observed at the boundaries of the successive outer zones.

The phase $\theta_\mathrm{d}$ is almost the same as $\theta_0$ outside the geometrical shadow. Inside the shadow, $\theta_\mathrm{d}$ goes behind as the receiver goes inside. The $\theta_\mathrm{d}$ changes rapidly in the vicinity of $A_\mathrm{d}$ minima. $\theta_\mathrm{d}$ on the axis differs by about $2\pi$ from $\theta_0$ with this setup. One cannot tell the difference by only observing phases for $\theta_\mathrm{d}$ and $\theta_0$ near the axis. However, the difference is clear by observing over a wide range of $r$. This is also seen in the vibration spiral as shown in Fig. 4 where $U_\mathrm{d}$ has rotated about $2\pi$.

The amplitude patterns for aperture $A_a$ and disk $A_\mathrm{d}$ are very different from each other as shown in Fig. 5 and one cannot tell the relation of $U_a$ and $U_\mathrm{d}$ by just looking at those patterns. However, Babinet's principle is satisfied when one compares the amplitudes and phase, off the axis. The phasors representing $U_0$, $U_a$ and $U_\mathrm{d}$ off the axis do not lie on the circle. Still they make a triangle as expected by Babinet's principle. However, it was found that the phases tend to coincide or to have a $\pi$ difference at 'characteristic' points. Three phases coincide just outside of the geometrical edge and $A_0$ is given by $A_a + A_\mathrm{d}$. The same relation is also observed at $\mathrm{M}_1$, although the error is large because $A_a$ is very small. At the maxima of $A_a$, $\theta_\mathrm{d}$ differs by $\pi$ from $\theta_0$ and $\theta_a$ and the patterns show $A_0 = A_a - A_\mathrm{d}$.

B. Phase difference

The amplitudes and phases obtained for the 12 mm circular aperture are plotted on a vibration spiral as shown in Fig. 6. The end points for phasor $U_{sa}$ are plotted with different symbols according to $z$. Arrow $U_{sa}$ shows the average value for all $z$ positions and $U_{sa3}$ shows that for $z = 38\lambda$. The phase difference $\varphi$ between the path via the center of the small aperture and via the edge was very small as shown in Table II. Only a tiny fraction of a half-period zone appears in the small aperture. The $\varphi$ values calculated by the path difference $\Delta_\mathrm{B}$ for $z = 17\lambda$ and $38\lambda$ are shown with bars on the arc. The observed $\varphi$ agreed well with the calculated values. However, the observed end points for $U_\mathrm{sa1}$ and $U_\mathrm{sa2}$ lay inside the circle as observed for $U_a$ in Experiment A; consequently, the values of $|\Delta\theta_\mathrm{sa}|$ are observed a little smaller than the theoretical values. The reason may be due to the finite size of the receiver and errors in alignment. Influences of these are expected to be larger for smaller $z$. On the other hand the points for $U_\mathrm{sa3}$ lay mostly on the arc and the $\Delta\theta$ value obtained was -85° and was close to the theoretical value of -88°. It is clearly seen that the wave from the center of the zone system is indeed $\pi/2$ ahead of the incident wave as expected from Huygens-Fresnel diffraction theory.

The radial diffraction patterns for the amplitudes $A_{sa1}$ and $A_{sa3}$ obtained for a small aperture at $z = 17\lambda$ and $38\lambda$, respectively, are shown in Fig. 7 together with $A_0$. Vertical lines represent the geometrical edges. A broad peak with a small structure at the center was observed for $A_{sa1}$. The amplitude on the axis is almost 1/10 that observed for the unobstructed wave. The pattern for $A_{sa3}$ is broader and the structure is not clear due to the smallness of the signal. The phases $\theta_{sa1}$ and $\theta_{sa3}$ observed at $z = 17\lambda$ and $38\lambda$, respectively, are also shown in Fig. 7. The dot-dash and broken parabolic curves are phase shift of $U_{sa}$, $2\pi(OR'-z)/\lambda + \Delta\theta$, calculated for the geometrical path difference between the center of the aperture to the receiver on the axis ($z$) and the center of the aperture to the receiver at $r$ ($OR'$), with an added phase difference $\Delta\theta$ of -74° or -90°. A value of -74° was chosen as the average value for $\theta_{sa}$ observed on the axis and -90° is the value for the small aperture limit. The measured phase agrees well with either of the curves. The result also shows that the unobstructed wave $U_0$ is indeed almost 90° behind the wave due to a very small aperture. The shape (width) of the parabolic curve is determined by the position of the origin, the center of the small aperture here, of the spherical wave.

## V. DISCUSSION

The measurement of phase in addition to the amplitude makes it possible to verify Babinet's principle directly in the Fresnel regime. The observation of phase also shows some characteristics of diffraction otherwise difficult to notice. The zone system has been studied and various applications have been made such as zone lenses in the X-ray region.[7,10,15] The present observation of a rapid phase change at an eclipse of a half-period zone shows another aspect of diffraction.

The phase and amplitude observed on the axis are simply predicted by the path difference between the wave along the axis and the wave via the edge of the aperture independent of the position of the aperture or the disk. The wave observed for a small aperture, which contains only a fraction of a half-period zone, has a phase always $\pi/2$ ahead of the wave without the aperture, provided that the aperture is small enough and is not too close to the source or the observation point.

The physical meaning of the origin of the $\pi/2$ phase difference is not fully understood. The presence of the factor -*i* in Eq. (2) gives the necessary π/2 phase difference. However, it indicates that the secondary sources oscillate a quarter of period ahead of the primary disturbance.[5,7] On the other hand, Huygens' principle states that the wavelets oscillate in phase with the primary wave. The parabolic curve $\theta_0$ in Fig. 7 represents a spherical wave centered at the transmitter, while $\theta_{sa1}$ or $\theta_{sa3}$, after passing through the small aperture, represent spherical waves centered at the small aperture. The latter can be regarded as an envelope of Huygens' spherical wavelet originated at the small aperture but with the phase shifted ahead by π/2.

A secondary wavelet approach by Huygens-Fresnel and an edge diffraction approach by Young are two fundamental physical models of diffraction[3,5] that could provide insight into the π/2 phase shift. The Huygens-Fresnel approach expresses the disturbance at *P* as the sum of all contributions from every point in the aperture as shown by the vibration spiral. The edge gives no contributions. An edge diffraction approach, the Maggi-Rubinowicz theory, describes the field behind the object as the superposition of the incident wave and the boundary diffraction wave ($U_B$, wave originated at the edge shifted by π) in the illuminated region, and, in the geometrical shadow, just as the boundary diffraction wave, this is. [3,5]

$$U_a(r) = A_0 \frac{\exp(ikr)}{r} + U_B(r) \qquad \text{when } P \text{ is in the direct beam,} \qquad (9)$$

$$U_a(r) = \qquad\qquad + U_B(r) \qquad \text{when } P \text{ is in the geometrical shadow.} \qquad (10)$$

Rubinowicz showed that the Kirchhoff diffraction integral (the mathematical expression of the Huygens-Fresnel approach) can be divided into the incident wave and the boundary wave, which is a line integral over the rim of an aperture, according to Young's approach.[3,5,16] The two approaches are considered as mathematically equivalent treatments. The edge diffraction theory, however has not been proved beyond the limits of validity of the Kirchhoff theory. ($z_0$, $z \gg D$, $D \gg \lambda$).[3,5] The physical reality of the boundary diffraction wave has been demonstrated by Ganci at the straight edge using the Young's double slit.[16]

We have tried to detect the boundary diffraction wave by observing the phase of the wave diffracted by a small aperture. The first terms in Eqs. (5) and (8) are the incident wave and the second terms ($-U_d$) can be regarded as the boundary diffraction wave $U_B$.[13,16] Then on the axis, we have $U_B = -U_d$ in Eqs. (5) and (8). As discussed in II.B., $U_0$ and $U_B$ have almost the same amplitude and a phase difference of π. Then, $U_a$ becomes π/2 ahead of $U_0$. The phase relation is also shown in the present measurements. When *P* moves deep into the shadow, $U_0$ vanishes and only $U_B$ remains. Then, $U_a$ should have a phase difference of $\pi$ in the shadow. However, the phase difference for $U_a$ remains $\pi/2$ both on the axis and in the geometrical shadow as shown by the parabolic curves $\theta_{sa1}$ and $\theta_{sa3}$ in Fig. 7. The reason may be that the "shadow" is not defined clearly when $D$ becomes small.[7] The present experimental setup is beyond the validity of the Kirchhoff theory ($z_0$, $z >> D >> \lambda$). Sophisticated measurements like Ganci's experiment[16] may be needed to observe the physical reality of the boundary diffraction wave.

The implemented apparatus was intended as a table-top instrument for the student laboratory and is far from ideal. However, it is still capable of showing fundamental aspects of the theory of diffraction. Ideally one needs a uniform wave and a huge room with no reflection in order to conduct a deeper investigation of the wave diffraction theory including the phase. In the classroom, however, satisfactory results can be obtained by setting the transmitter and the receiver close to each other with a proper alignment. Since Babinet's principle can be studied directly, the diffracting object need not be circular. An annular shape,[14,15] an elliptical form or squares, in fact any form can be used.

## VI. CONCLUSIONS

Babinet's principle in the extreme Fresnel regime has been directly verified by observing ultrasound waves diffracted by a circular aperture and a complementary disk. The phase as well as the amplitude were measured and analyzed by a graphical treatment. It has been also found that the wave without the diffracting objects is indeed $\pi/2$ behind the wave from the center of the zone system, which has been regarded as a defect of Fresnel's theory. The $\pi/2$ phase difference appears also in Fresnel-Kirchhoff diffraction, Rayleigh-Sommerfeld diffraction and even in the edge-diffracted approach by Young.

## ACKNOWLEDEMENTS

We are also grateful to Prof. T. A. King of the University of Manchester for reading of the manuscript and for his valuable comments. The authors would like to thank researchers in physics of Kitasato University for a valuable discussion.

Table I. Experimental parameters for the study of Babinet's principle. Values for $A_d/A_0$ are obtained by closed solutions, Eqs. (5) and (6), for Rayleigh-Sommerfeld diffraction theory.[12]

| | | | | |
|---|---|---|---|---|
| frequency | | 40.0 kHz | | |
| wavelength λ at room temperature | | 8.6 mm | | |
| $D$ | | 10 λ | | |
| $z_0 + z$ | | 100 λ | | |
| $z$ | (λ) | 17 | 26.3 | 38 |
| $\Delta_B$ | (λ) | 0.87 | 0.64 | 0.53 |
| φ | (°) | 313 | 230 | 190 |
| $A_d/A_0$ | | 0.95 | 0.98 | 0.99 |

Table II. Experimental parameters for the study of the vibration spiral. Δθ is the phase difference from the incident wave.

| | | | | |
|---|---|---|---|---|
| frequency | | 40.0 kHz | | |
| wavelength λ at room temperature | | 8.6 mm | | |
| $D$ | | 1.4 λ | | |
| $z_0 + z$ | | 100 λ | | |
| sublabels | | sa1 | sa2 | sa3 |
| $z$ | (λ) | 17 | 26.3 | 38 |
| $\Delta_B$ | (λ) | 0.017 | 0.013 | 0.010 |
| φ | (°) | 6.3 | 4.6 | 3.7 |
| Δθ (=-90+φ/2) | (°) | -87 | -88 | -88 |
| $A_d/A_0$ | | 1.00 | 1.00 | 1.00 |

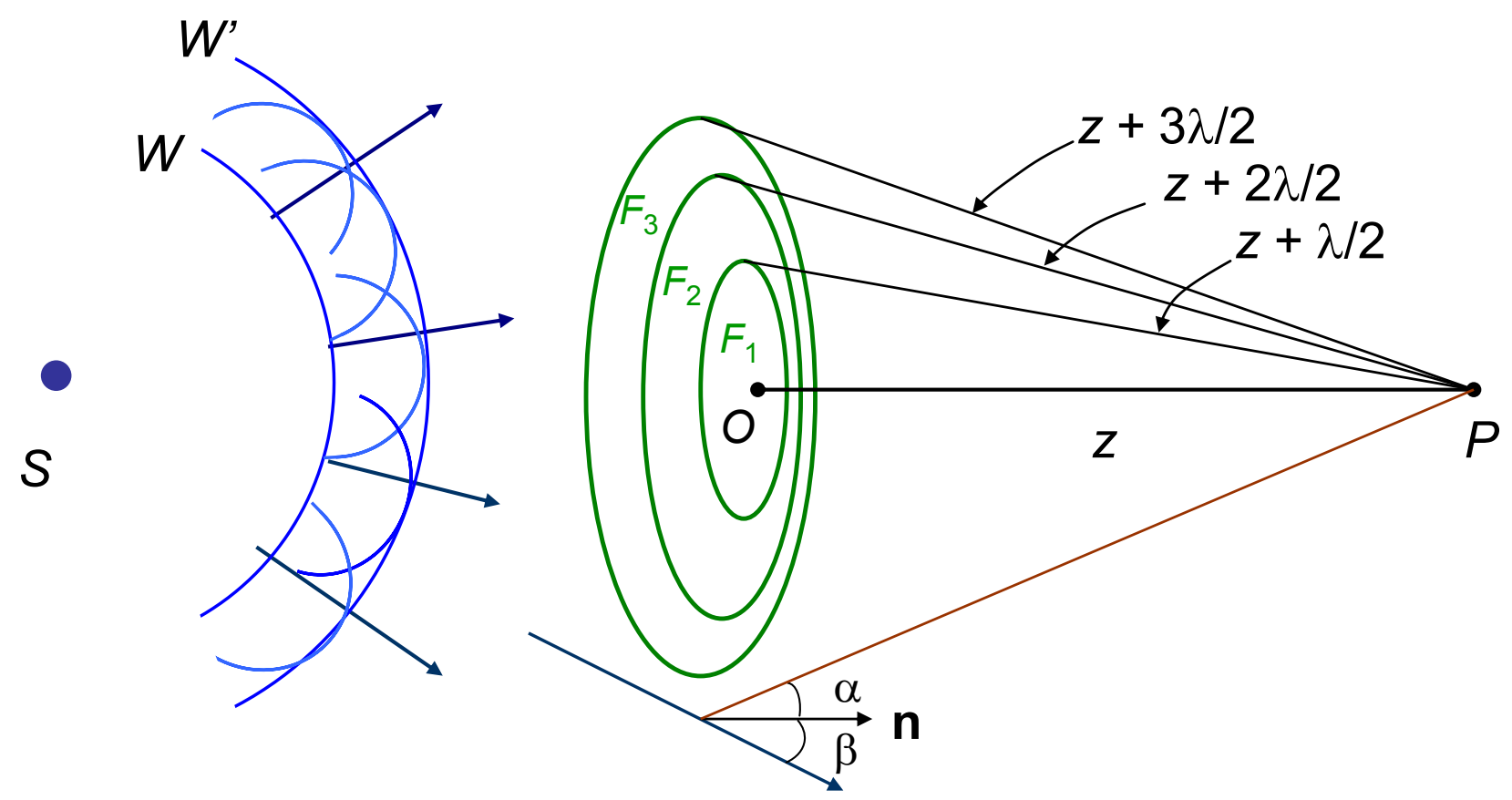

Fig. 1. Huygens' principle and construction of Fresnel's half-period zones ($F_1$, $F_2$, $F_3$) on a spherical wave front at a circular aperture. A spherical wave front $W$ is originated at $S$. $O$ is the center of the Fresnel zone and $P$ is the observation point. Huygens' secondary wavelets originating on $W$ combine to form a new wave front $W'$. **n** is a normal unit vector perpendicular to the diffraction plane.

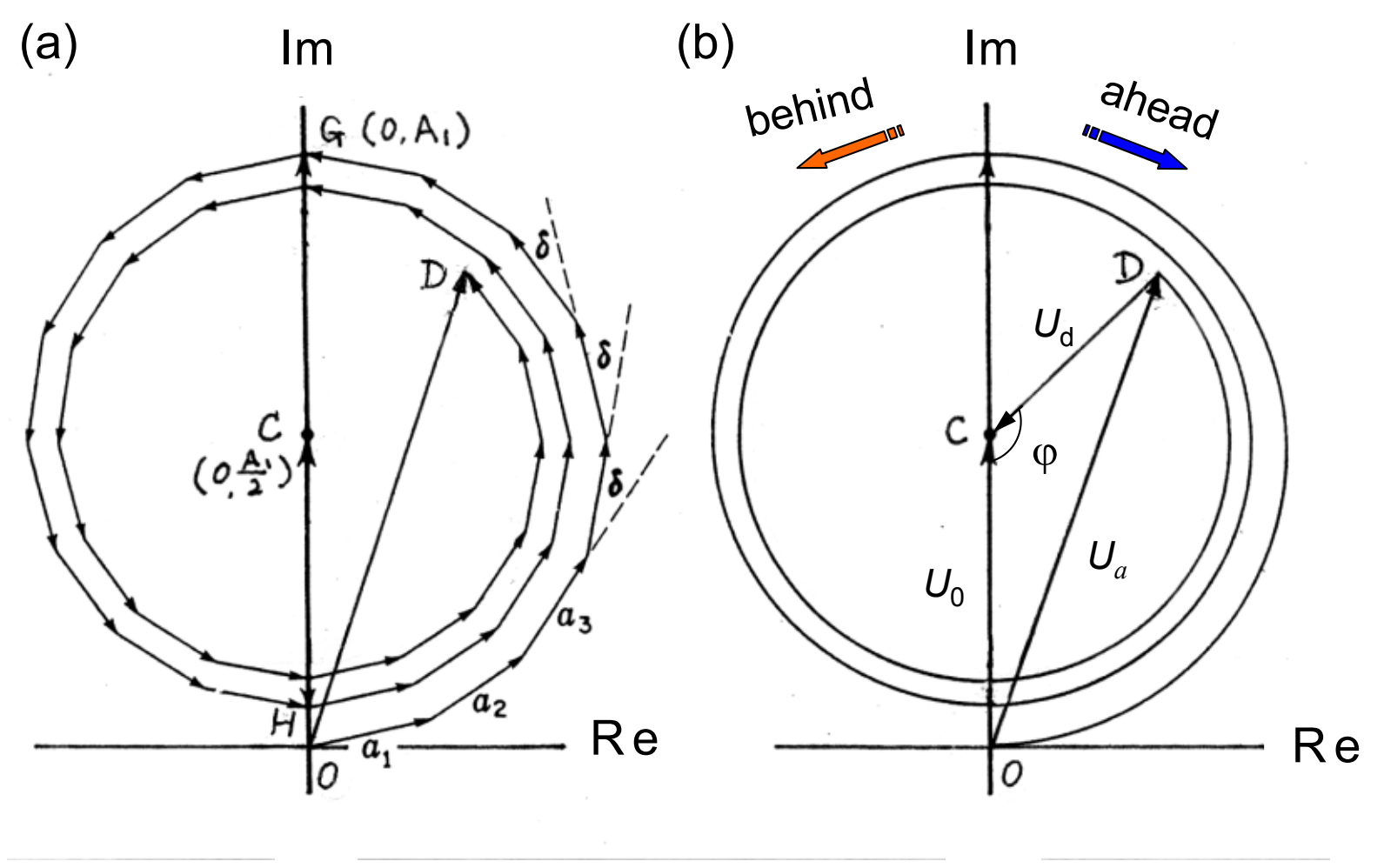

Fig. 2. Vibration spiral for Fresnel half-period zones of a circular aperture. (a) half-period zones are divided into 8 subzones, (b) Continuous subdivision of the zones.

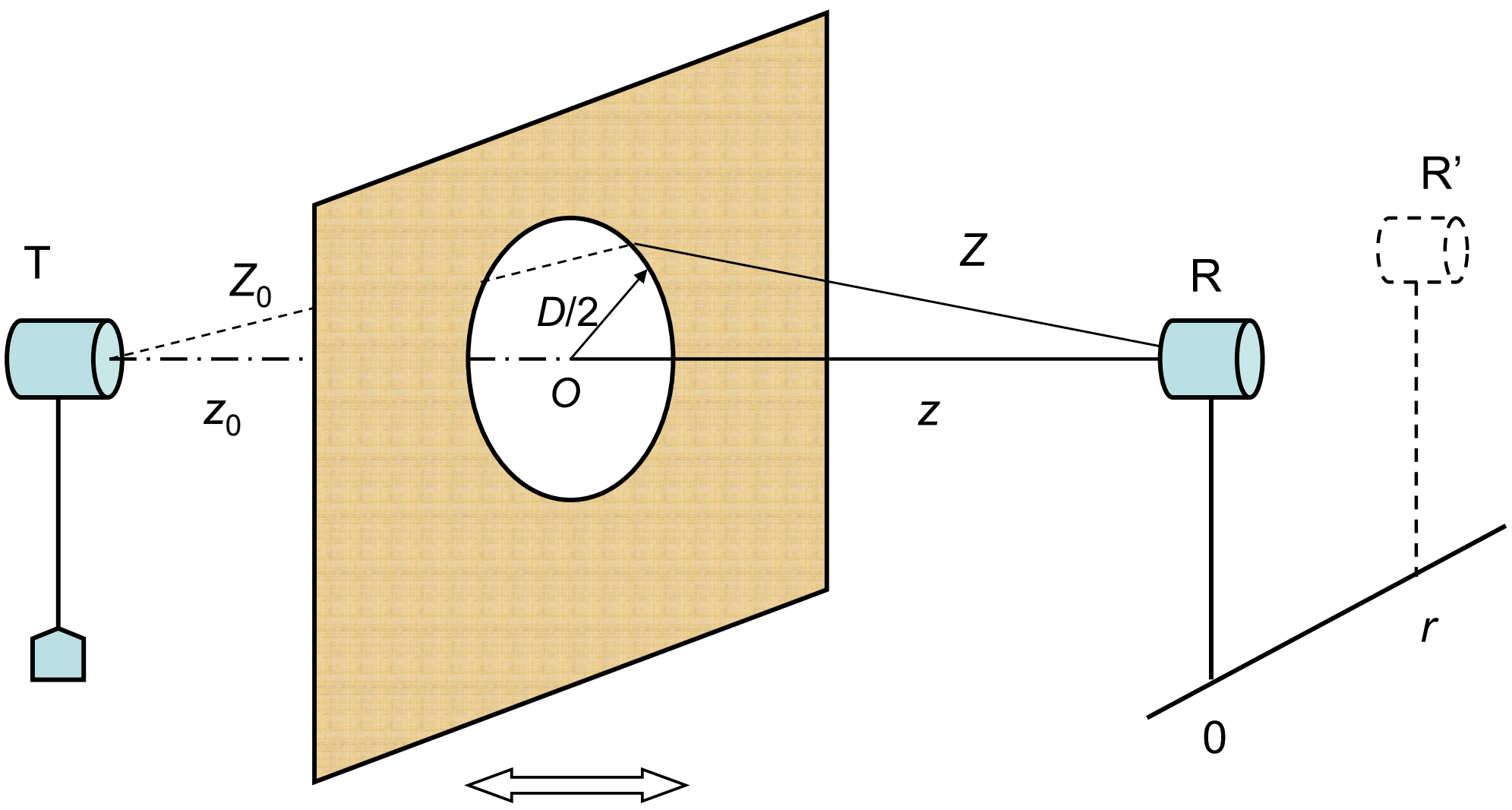

Fig. 3. Schematic view of the experimental setup with a circular aperture. The data for the vibration spiral (Fig. 4) were taken on the axis ($r = 0$) by changing $z$.

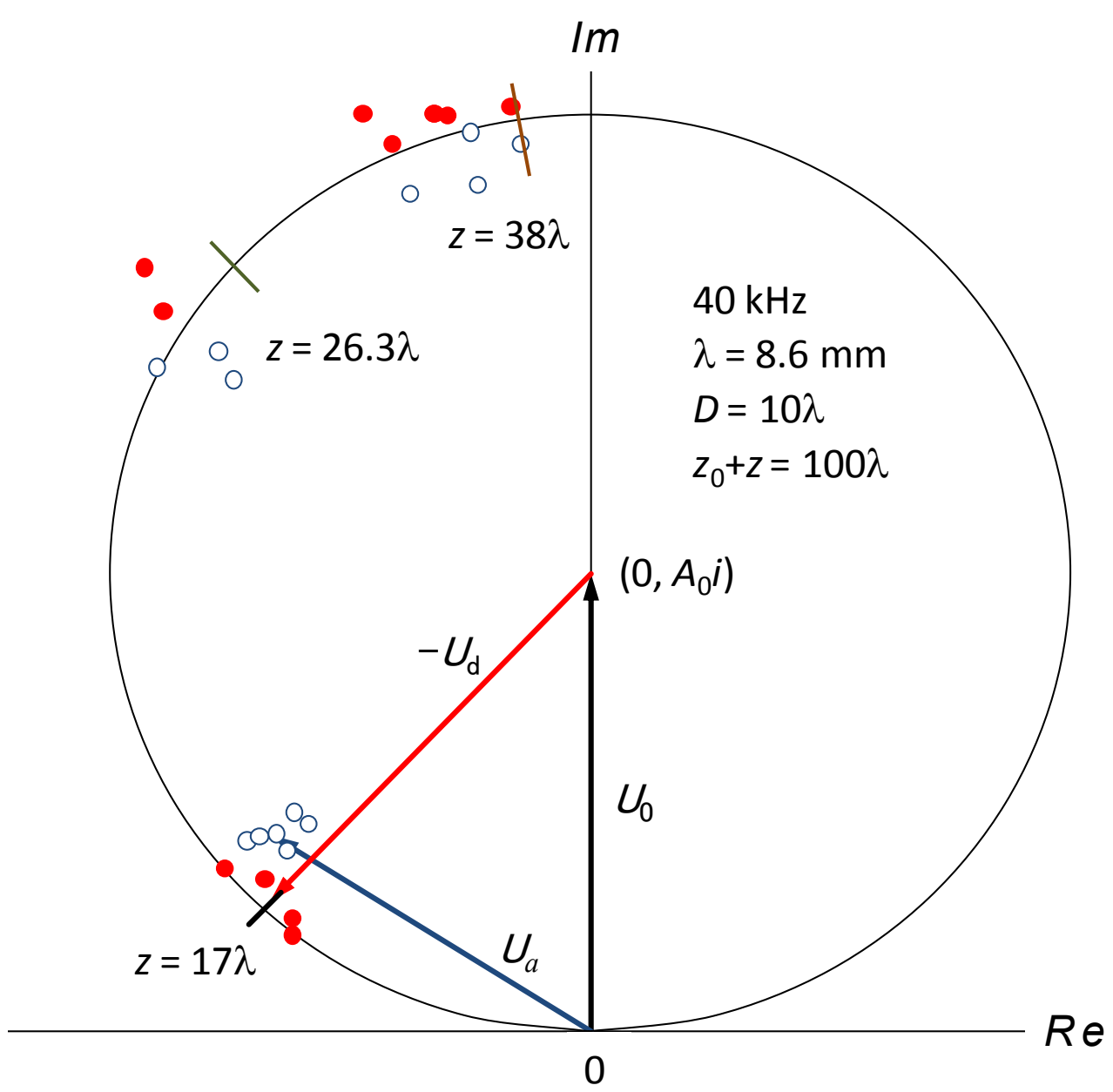

Fig. 4. The amplitude and phase relation obtained for unobstructed wave $U_0$, a circular aperture $U_a$ and a circular disk $U_d$. The receiver was set on the axis. The phase $\theta_0$ for $U_0$ was taken to be $\pi/2$. The end points for phasors $U_a$ and $U_0$ - $U_d$ are plotted with open and closed symbols, respectively. The $\varphi$ values calculated for each setup are shown with bars on the circle.

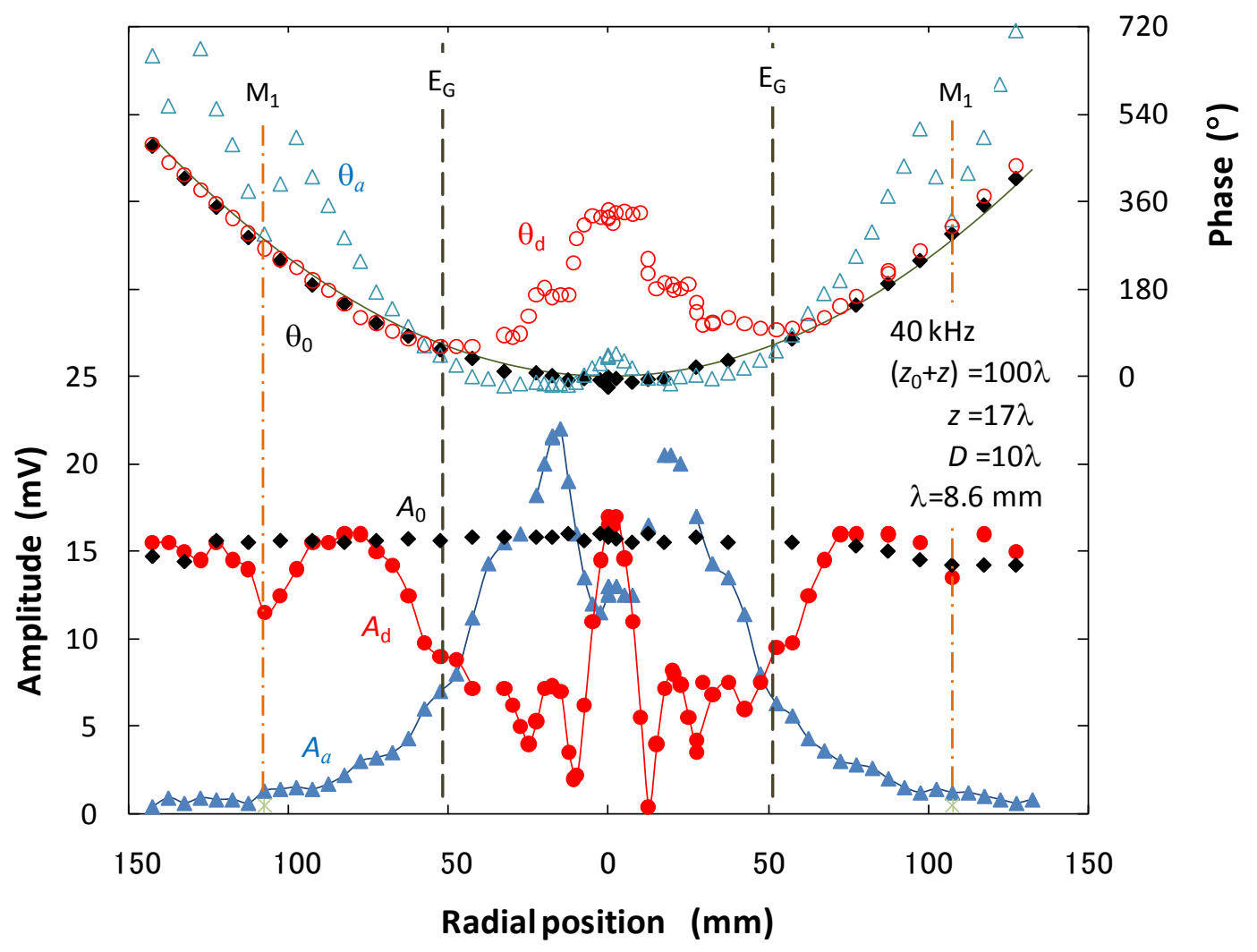

Fig. 5. The diffraction patterns for the amplitude $A$ (bottom) and the phase θ (top) as a function of the radial distance $r$ from the axis observed for complementary circular aperture and disk. The vertical broken lines ($E_G$) show the geometrical edges and dot-dash lines ($M_1$) represent positions for the eclipse of the first half-period zone (see the text). The phase range goes from below 0° up to 720° because of the continuity of the wave.

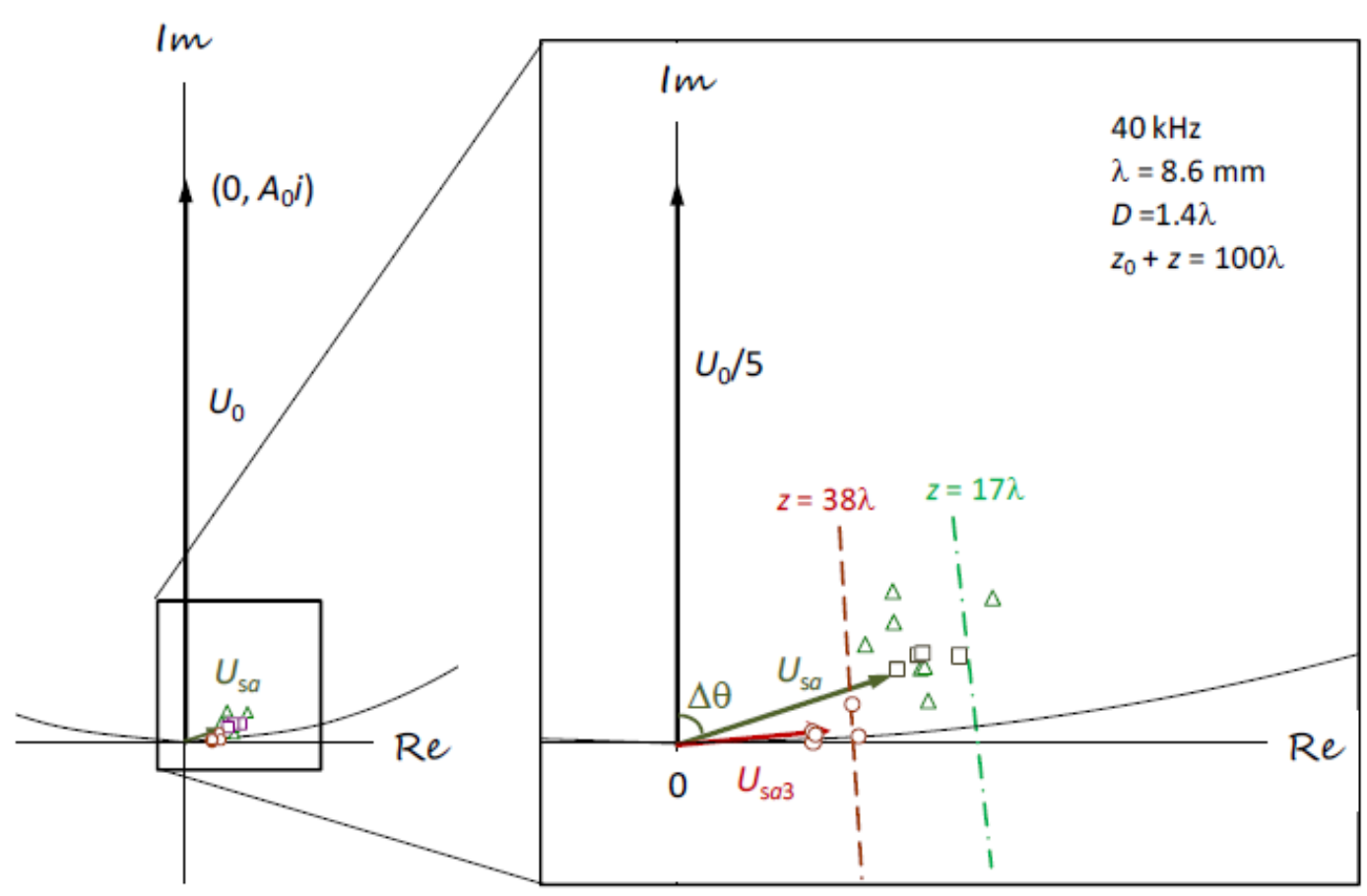

Fig. 6. The amplitude and phase relation obtained for a small aperture of 1.4λ in diameter. The receiver was set on the axis. The area enclosed by a square is displayed on an expanded scale on the right. The end points for phasors $U_{sa1}$, $U_{sa2}$, and $U_{sa3}$, at distance $z$ of 17λ, 26.3λ and 38λ, are plotted with triangles, squares and circles, respectively. Arrow $U_{sa}$ shows the average value for all $z$ positions and $U_{sa3}$ shows that for $z$ = 38λ. The φ values calculated for $z$ = 17λ and 38λ are indicated with bars on the arc.

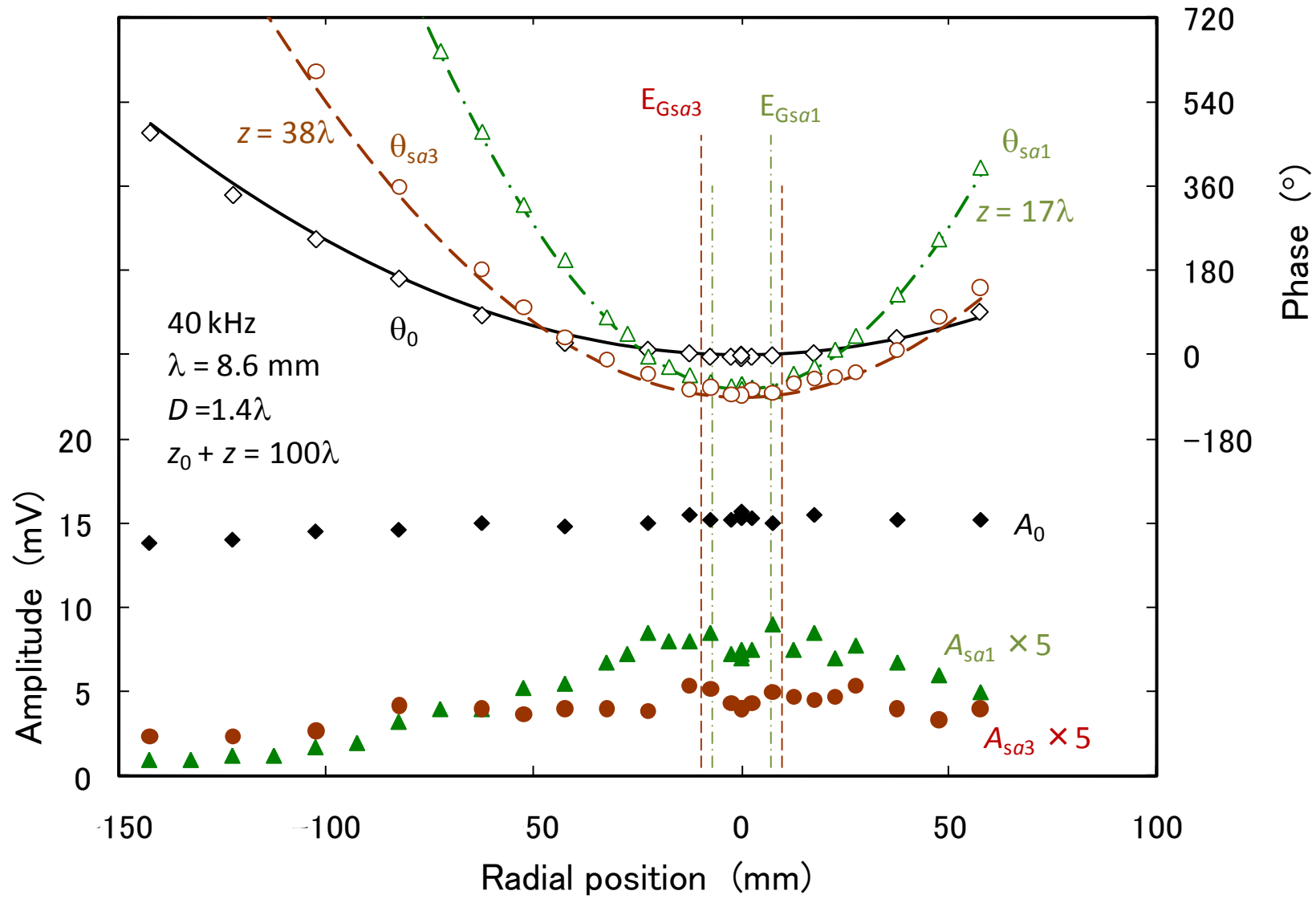

Fig. 7. The diffraction patterns for the amplitude $A$ (bottom) and the phase θ (top) as a function of the radial distance observed for a small aperture of 1.4λ in diameter. Sublabels 1 and 3 stand for the distances $z$ of 17λ and 38λ, respectively. Parabolic curves show phase differences calculated for the geometrical path difference between the center of the small aperture to the point on the axis ($z$) and the center of the aperture to the receiver ($OR$') with an added phase difference of -74° or -90° (see the text). The vertical lines ($E_{Gsa}$) show the geometrical edges.